# Interplay among Work Function, electronic structure and stoichiometry in nanostructured vanadium oxides films

*Alessandro D'Elia[a,b,*], Cinzia Cepek[b], Monica de Simone[b], Salvatore Macis[c], Blaž Belec[d], Mattia Fanetti[d], Paolo Piseri[e], Augusto Marcelli[f,g,h], Marcello Coreno[h,i]*

a) Department of Physics, University of Trieste, Via A. Valerio 2, 34127 Trieste, Italy

b) IOM-CNR, Laboratorio TASC, Basovizza SS-14, km 163.5, 34012 Trieste, Italy;

c) Department of Physics, Università Sapienza, Piazzale Aldo Moro 5, 00185 Rome, Italy

d) Materials Research Laboratory, University of Nova Gorica, Vipavska 13, 5000 Nova Gorica, Slovenia;

e) Dipartimento di Fisica & CIMaINa, Università degli Studi di Milano, via Celoria 16, 20133 Milano, Italy;

f) Istituto Nazionale di Fisica Nucleare, Laboratori Nazionali di Frascati, 00044 Frascati, Italy;

g) Rome International Centre for Material Science Superstripes, RICMASS, Via dei Sabelli 119A, 00185 Rome, Italy

h) ISM-CNR, Istituto Struttura della Materia, LD2 Unit, Basovizza Area Science Park, 34149 Trieste, Italy

i) Elettra-Sincrotrone Trieste, Basovizza 34149, Italy;

\* **Correspondence**: delia@iom.cnr.it;

**Abstract:** The work function is the parameter of greatest interest in many technological applications involving charge exchange mechanisms at the interface. The possibility to produce samples with a controlled work function is then particularly interesting, albeit challenging. We synthetized nanostructured vanadium oxides films by a room temperature Supersonic Cluster Beam Deposition method, obtaining samples with tunable stoichiometry and work function (3.7-7 eV). We present an investigation of the electronic structure of several vanadium oxides films as a function of the oxygen content via in-situ Auger, valence-band photoemission spectroscopy and work function measurements. The experiments probed the partial $3d$ density of states, highlighting the presence of strong V$3d$-O$2p$ and V$3d$-V$4s$ hybridization which influence $3d$ occupation. We show how controlling the stoichiometry of the sample implies a control over work function, and that the access to nanoscale quantum confinement can be exploited to increase the work function of the sample relative to the bulk analogue. In general, the knowledge of the interplay among work function, electronic structure, and stoichiometry is strategic to match nanostructured oxides to their target applications.

# Introduction

At present, the demand of energy is one of the main concerns of all major economies, driving a strong worldwide demand for low-power-consumption electronics. The development of a technology easy to handle on large scale and able to provide efficient devices is of primary importance for our society. The interest in organic semiconductors, has risen strongly because of high efficiency applications (*e.g.* OFET, OPV and OLED)[1,2] and for the eases of manipulation at industrial level.[3] Organic semiconductors do not have intrinsic charge carriers: these have to be supplied by electrodes, *i.e.*, hole and electron injection layers. Therefore, a typical organic semiconductor device (OSD), consists of an active layer sandwitched between two electrodes. The optimization of the charge injection at the electrode-active layer interface is paramount to maximize the efficiency of the device.[2,4,5]

In general, this can be accomplished by tailoring the energy levels of the organic semiconductor (HOMO, LUMO and Ionization Potential, IP) and the work function (WF) of the electrodes.[6] In the active layer, the HOMO and LUMO act respectively as donor and acceptor levels while in the electrodes the Fermi level (FL) plays both roles. Ideally to minimize energy losses, the FL of Hole Injection Layers in OSDs should be as close as possible to the HOMO level (high WF) of the active layer and the FL of the electron injection layer should be close to the LUMO level (low WF).[6–8] Therefore, control over the electrodes WF is highly desirable for this application. Transition metal oxides (TMO) have been investigated because of their versatility, since their electrical and chemical properties can be tuned to maximize the charge injection within molecular interfaces.[4–6,8,9] TMOs have been used as efficient Hole Injection Layers (HIL) ($MoO_3$, $WO_3$, $V_2O_5$ [8,10,11]) and electron injection layer ($ZrO_2$ [12]).

Among TMOs, vanadium oxides are interesting materials for technological applications. Vanadium ([Ar] $3d^3 4s^2$) is a very reactive element and different oxides such as VO, $V_2O_3$, $VO_2$ and $V_2O_5$ characterized by the oxidation states: $V^{+2}$, $V^{+3}$, $V^{+4}$ and $V^{+5}$, respectively, can be synthetized. The WF of vanadium oxides can be tuned by changing the stoichiometry of the sample,[6,13,14] or its dimensionality. The nanoscale dimensions in thin films and nanoparticles can be an important degree of freedom to control the WF of any oxide, resulting in significant deviations of the electronic structure from that of the bulk. For nanoparticles the reduced size enhances the electronic density confinement leading to a clear size dependence of the WF.[15–18] Moreover, oxides are prime cases of strong-electronic-correlation systems. $VO_2$ and $V_2O_3$ exhibit sharp, fast and reversible metal insulator transitions (MIT) triggered by temperature[19] and characterized by a nanoscale phase separation.[20–23] Synthetizing nano-sized samples with a tunable electronic structure is therefore relevant for many applications.

Here we show how it is possible to control the stoichiometry of nanostructured (NS) vanadium oxide films using a pulsed-vaporization cluster source (namely the Pulsed Microplasma Cluster Source – PMCS [24–26]) and the Supersonic Cluster Beam Deposition (SCBD) method. Synthetizing samples with tunable oxidation state allowed us to unravel the systematic changes of the electronic structure of vanadium oxides NS films as a function of oxygen content. Understanding the evolution of the valence band (VB) electronic structure as a function of stoichiometry allows tailoring the chemical reactivity of a sample. An accurate knowledge of the interplay between valence band features, WF, and stoichiometry, is essential to match the optimum condition for a desired application: it is particularly important in OSD where energy level alignment plays a crucial role in determining the efficiency of the device.

## Results and discussion

### Nanoparticles size and morphology

In order to investigate the morphology of nanoparticles deposited by SCBD, a deposition of 0.3 sec has been performed on a TEM grid with amorphous C supporting film. In fig.1 the TEM image is shown for $VO_{2.2}$. Even though the nanoparticles aggregate into nanoparticle agglomerates, it is still possible to clearly distinguish them. The intense agglomeration of nanoparticles even after few second deposition makes TEM images unsuitable for accurate size distribution analysis. Still, the size of the nanoparticles can be assessed in the range 3-6 nm for most of them. No large differences have been observed for samples with different deposition parameters. Preliminary in-situ XANES study showed that nanostructured films are actually amorphous (See supplementary materials). However, in high magnification TEM images a crystalline lattice is in some cases observed (see Fig.S11 in Supplementary), which is ascribed to crystallization of amorphous material under the high energy TEM electron beam.

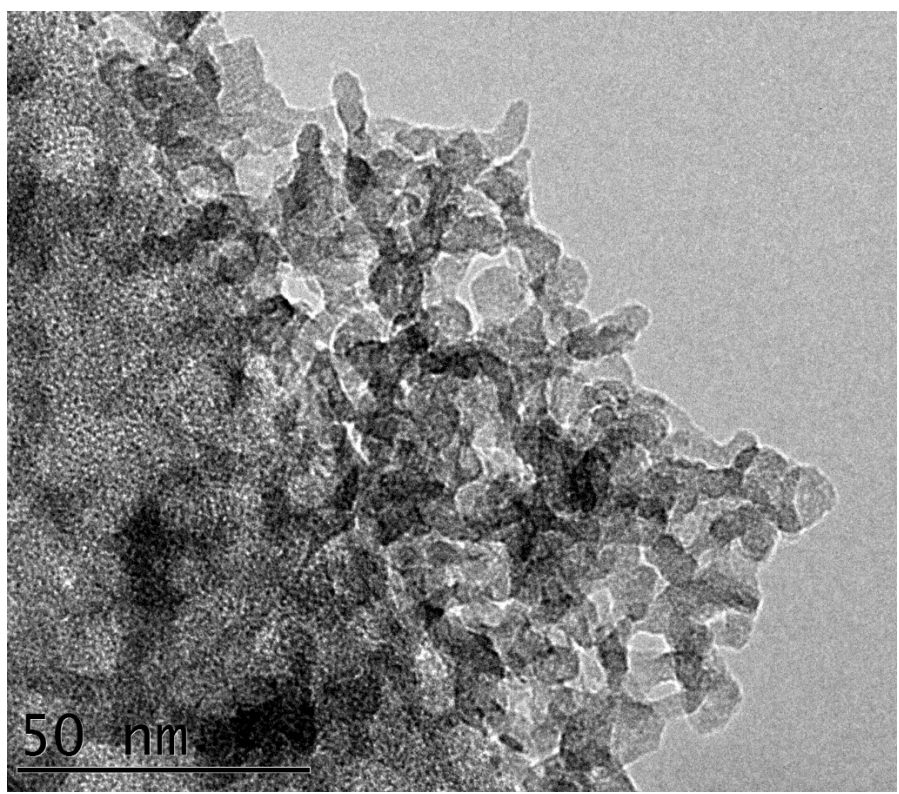

**Figure 1:** TEM image of the nanoparticle assembled nanostructured film for $VO_{2.2}$

## Oxidation control and measuring the stoichiometry of the films

To investigate the stoichiometry, *i.e.*, the ratio between oxygen and vanadium atoms, we performed core-level XPS spectra of V *2p* and O *1s* electrons. Quantitative information has been extracted by fitting simultaneously the line-shape of V *2p* and O *1s* core levels which is a known indicator of the stoichiometry.[27] For each component (vanadium and oxygen) we used a pseudo Voigt curve;[28] we report the complete set of fitting parameters in the supplementary material.

From now on we use the stoichiometric ratio thus measured, $x=$ [# of oxygen atoms]/[# of vanadium atoms], in the range $0 \leq x \leq 2.5$ to identify the samples since exists a one to one correspondence between x and a specific sample.

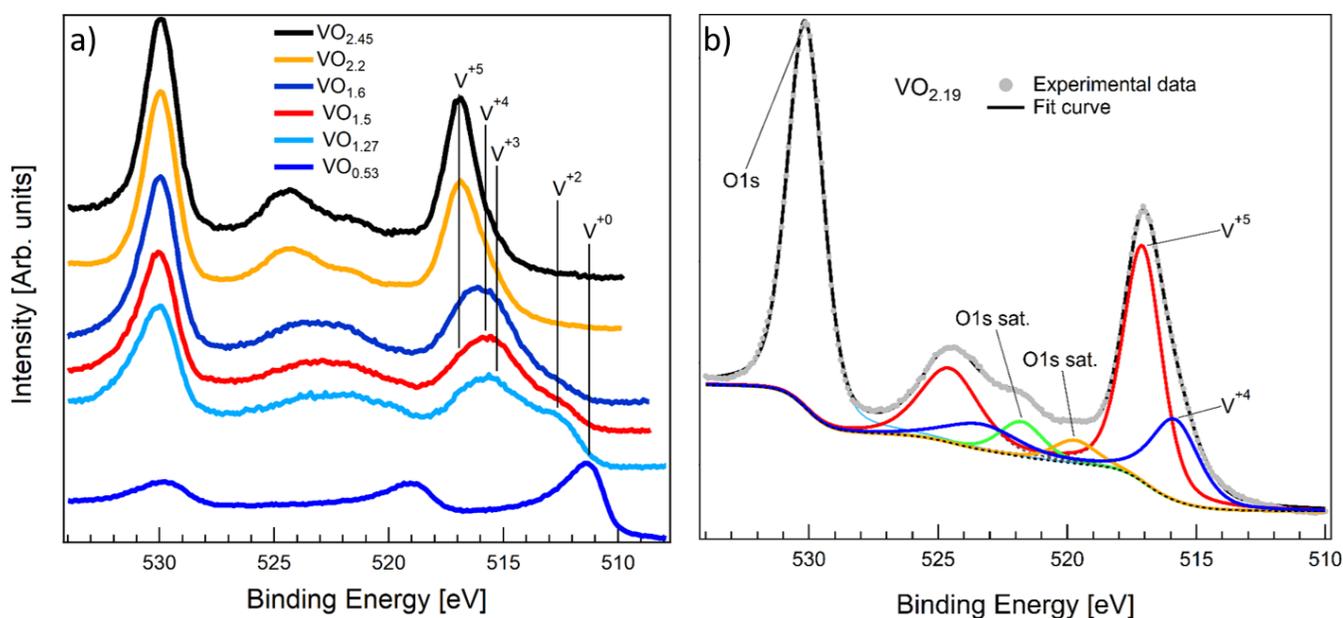

**Figure 2:** V2p and O1s spectra of vanadium and oxygen. a) V2p and O1s core level spectra of all samples. Peak positions of the vanadium components for different oxidation state are highlighted by straight vertical lines; b) V2p and O1s core level spectra and fit curve of $VO_{2.2}$. The individual fit components are shown as coloured curves. The label O 1s sat indicates the features generated by the Mg Kα3 and Mg Kα4 component of the radiation.

The core level spectra of V *2p* and O *1s* electrons of different vanadium oxides synthesized by SCBD are shown in the left panel in Fig. 2; the right panel shows the decomposition of the experimental spectrum into its different fit components for a selected sample ($VO_{2.2}$). This result proves that the applied method may be used to synthesize $VO_x$ films with controllable stoichiometry in the range $0.5<x<2.45$. The complete set of core level spectra used in this work can be found in the supplementary materials.

Films with stoichiometry $x<1$ ($VO_{0.53}$ in Figure 2) are called sub-oxides, *i.e.*, oxides in which the electropositive element is more abundant than oxygen. Generally, only a severe temperature treatment (T>1000 K) under high vacuum allows obtaining vanadium sub-oxides [29,30] while using SCBD method we are able to synthetize them at room temperature.

**Auger $L_3M_{2,3}M_{4,5}$ spectroscopy**

The stoichiometric vanadium oxides: VO, $V_2O_3$, $VO_2$ and $V_2O_5$ have nominally 3, 2, 1 and 0 *3d* unpaired electrons, respectively. Auger decays with one hole in the VB (*i.e.*, of the form V $XYM_{4,5}$) probe the V *3d* partial density of states (DOS) and therefore can be linked to *3d* occupation number and to sample stoichiometry. The atomic vanadium electronic configuration is [Ar] $3d^3 4s^2$, so that the Auger channel involves *3d* electrons while the 4s electron contribution is negligible.

The ideal Auger decay to probe the partial *3d* DOS is the V $L_3M_{4,5}M_{4,5,}$ characterized by two holes in the conduction band. Unfortunately, in vanadium oxides this decay channel is obscured by the O K $L_{2,3}L_{2,3}$ Auger electrons, even for an extremely low amount of oxygen.[31] V $L_3M_{2,3}M_{4,5}$ Auger decay is experimentally observable and *3d* electrons give the main contribution to this channel.[32,33]

As shown in Fig. 3a the Auger $L_3M_{2,3}M_{4,5}$ of vanadium oxides exhibits two main features labelled V and U. As pointed out by Sawatzky and Post, these two components are correlated with the oxidation state.[34] Feature V becomes dominant as the oxidation state increases: it is related to V *3d* electrons covalently bound to O *2p* electrons. Feature U is associated to unpaired *3d* electrons. To extract quantitative information about the *3d* contribution in the VB, we fit the Auger V $L_3M_{2,3}M_{4,5}$ spectra using three pseudo Voigt curves: one for feature V, one for feature U, and the last for the small contribution from the O K $L_1L_1$ decay at ~477 eV.

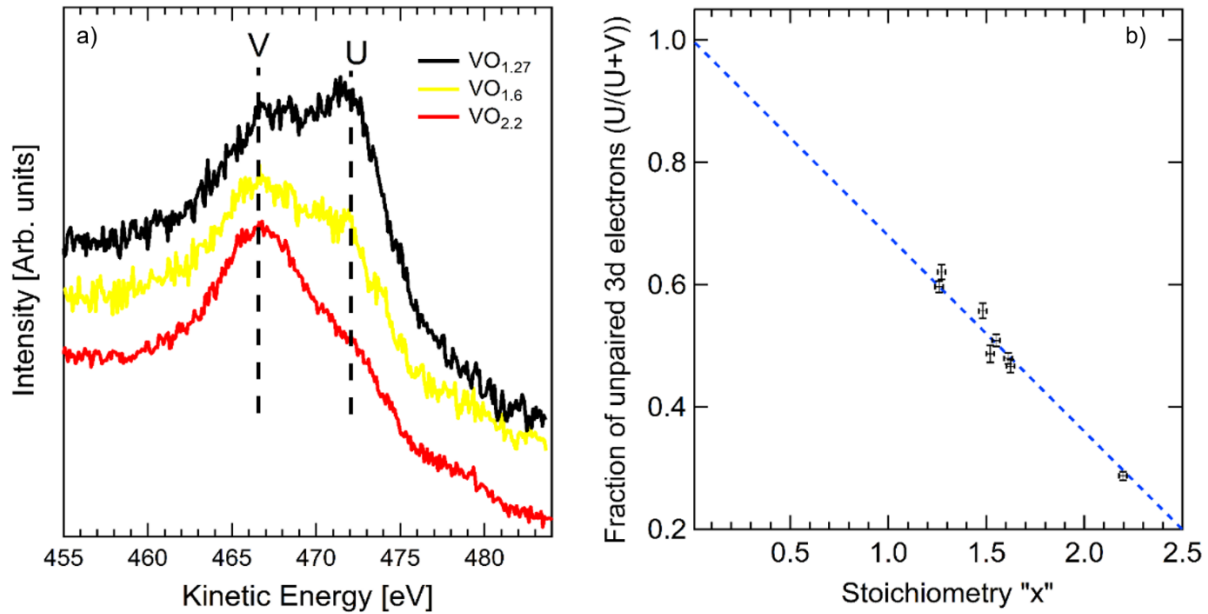

**Figure 3:** a) Comparison of Auger L3M2,3M4,5 spectra of vanadium oxides NS films with stoichiometric ratio 1.27, 1.6 and 2.2 in the kinetic energy range 455-482 eV. The spectra are vertically shifted for clarity in order to highlight the behavior of features labelled U and V (see text). b) Correlation between stoichiometry and Auger fraction (U/(U+V)) as obtained by fitting the Auger L-3M2,3M4,5 spectra. The blue line is the best least-squares fit of the experimental data with a straight line of fixed intercept (at x=0 the fraction of unpaired electrons must be equal to 1). The best-fit slope is k = -0.320±0.005.

The background has been modelled using the Shirley curve. The Auger fraction (*U/(U+V)*) where *U* and *V* refer to the area of the two features), can be associated to the fraction of unpaired *3d* electrons relative to the total number of *3d* electrons. The behaviour of the Auger fraction vs. *x* is reported in Fig. 3b, along with the best fit to a straight line (intercept =*1*, best fit slope is k =*-0.320±0.005*).

In the pure metal, no oxygen atoms are available to form V-O bonds, thus $V = 0$ and $U/(U+V)=1$; accordingly, in the straight-line fit the intercept is held at $1$. Likewise, in the maximum oxidation state ($x=2.5$) the fraction should be zero because no unpaired $3d$ electrons are present and U=0. Actually, the extrapolated value for $x=2.5$ is different from zero: $U/(U+V)|_{2.5} =0.2\pm0.01$, denoting a partial $3d$ occupancy. This is not unusual for $3d^0$ compounds because of the strong $3d$-$2p$ hybridization in the specific case of $V_2O_5$, these results confirm previous published resonant photoemission (*ResPe*s) experiments.[35,36] The number of unpaired $3d$ electrons in the samples has been calculated multiplying by 3 (which is the number of the nominal $3d$ electrons in vanadium atom) the Auger fraction $U/(U+V)$. For the stoichiometry value for which no experimental data are available, the Auger fraction value has been extrapolated by the straight line fit of the experimental data. In Table 2 we report the number of unpaired $3d$ electrons per vanadium atom equal under the above assumptions. They are not in good agreement with the occupation number calculated by Zimmerman and co-workers,[37] suggesting that further theoretical investigations are necessary.

We would like to underline here also that data in Table 1 refer to the core-hole ionized systems and do not necessarily reflect neutral $VO_x$ features. For example, on the one hand, the extrapolated values are in good agreement with measured values of unpaired $3d$ electrons for $VO_2$, suggesting that the core-hole effect in this oxide is negligible. On the other hand, the data for the $V_2O_3$ oxide suggest that only 1.56, instead of two, $3d$ electrons are unpaired. In this system, we point out the presence of a mixed band of $4s$-$3d$ character near the Fermi energy.

**Table 1:** Extrapolated values of the unpaired $3d$ electrons per vanadium atom.

| Stoichiometric oxides (oxidation state) | # of unpaired $3d$ electrons per V atoms (this work) | # of unpaired $3d$ electrons per V atoms (Zimmerman et al.) | Nominal # of unpaired $3d$ electrons per V atom |
|---|---|---|---|
| $V_2O_5$ (+5) | $0.6 \pm 0.04$ | 1.3 | 0 |
| $VO_2$ (+4) | $1.08 \pm 0.03$ | 1.9 | 1 |
| V₂O₃ (+3) | 1.56 ± 0.02 | 2.6 | 2 |
| VO (+2) | 2.72 ± 0.02 | - | 3 |

However, for the $V_2O_3$ these results are not in agreement with Hard-X-Ray PES spectra which showed a VB with a pure *3d* character near the Fermi level and 4s character centred at ~8 eV binding energy.[38] This suggests us that in $V_2O_3$ the core-hole effect is relevant. For VO no data are available in the literature but a strong 4s-3d hybridization is known to occur. To reproduce the VB features of this system using LCAO (Linear combination of atomic orbitals) and APW (Augmented plan-wave) computational methods, the electron configuration {V($3d^4 4s^1$) O($2p^4$)}, which implies a strong *4s-3d* hybridization,[39–41] has to be considered. Moreover, in this case, the percentage of unpaired *3d* electrons has to be multiplied by four instead of three, obtaining the value of 2.72 ± 0.02 unpaired electrons.

**Valence band PES**

Valence band spectra of vanadium oxides can be divided into two regions along the Binding Energy (BE) axis as can be seen in Fig. 4a: the O *2p* band (2.5 - 10 eV BE, Fig. 4b) and the V *3d* (Fermi level - 2.5 eV BE, Fig. 4c)

The V *3d* region is populated by unpaired *3d* electrons. The corresponding spectral feature shape and position depend on the oxygen amount present in the material, which determines the DOS via crystal field effects. The position of the V *3d* band exposes the metal character of sub-oxides, *i.e.*, a finite DOS at the Fermi level. This indicates that the amount of oxygen in these samples is not sufficient to break the degeneracy of *3d* levels [6] and that the independent particle approximation is sufficient for a theoretical band description. V *3d* band of $VO_{1.27}$ clearly exhibits two features, one centred around 1.5 eV and a second at ~0.4 eV (indicated by arrow in Fig. 4c. These features can be associated to the $V^{3+}$ and $V^{2+}$ components of the core-level spectra (See supplementary materials). A small, but finite DOS at the Fermi level is also present, supporting the picture of vanishing of the *3d* degeneracy. For higher values of the stoichiometric ratio, we observe a decrease of the V *3d* band spectral weight, which is transferred to the O *2p* band. An interesting case is that of $VO_{2.45}$, a defective variant of $V_2O_5$, whose V *3d* band is populated due to oxygen vacancies. The latter generates $V^{4+}$ ions, because of the large *3d-2p* hybridization.[34,42]

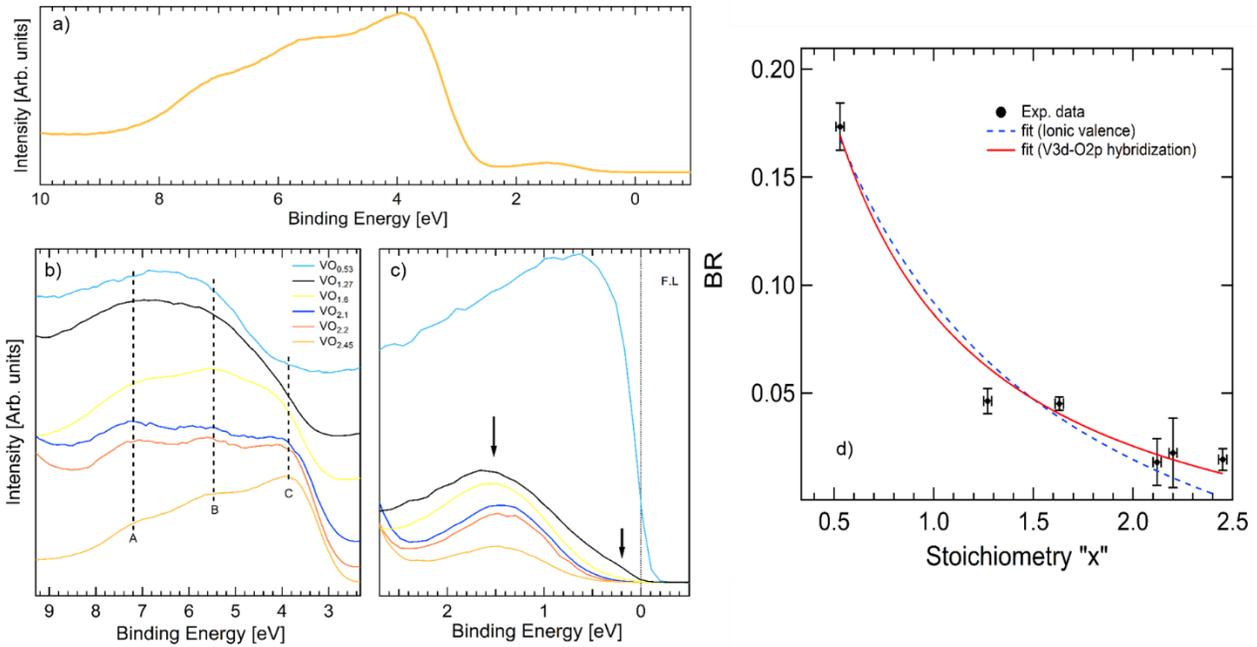

**Figure 4:** a) full valence band spectrum of $VO_{2.45}$ in the BE region [0-10] eV reported as representative spectra. O 2p and V 3d regions of the valence band spectra (b and c respectively) for $VO_{0.53}$, $VO_{1.27}$, $VO_{1.6}$, $VO_{2.1}$, $VO_{2.2}$ and $VO_{2.45}$ oxides: b) Comparison of the spectra of the O 2p region. Spectra are vertically shifted in order to highlight the behavior of the feature labelled A, B and C; c) comparison of the spectra in the V 3d region, near the Fermi level (dotted line). The arrows indicate $VO_{1.6}$ features. The intensity of the spectra is normalized to the maximum value of O2p region. d) BR as a function of the stoichiometric ratio. The red curve is the fit obtained using the eq. 1 including the 3d-2p hybridization term; the blue dotted curve is obtained considering ionic model of the valence band.

As shown in Fig. 4b, the O *2p* region exhibits a fine structure due to the superposition of V *4s*, V *3d* and O *2p* orbitals. To qualitatively describe these VB features we use the LCAO band model terminology. For the high oxidation states, three features (labelled A, B and C) are clearly visible, whose relative intensity evolves with the stoichiometry. The A component is assigned to the superposition of V 4s and O *2p* orbitals.[38] The DOS behind the B feature is dominated by the mixture of O *2p* – and V *3d* orbitals, while the electron yield in the C region is dominated by the non-bonding O π* orbitals with a small contribution from V *3d* electrons.[43,44] For the samples with lower oxidation, the C feature decreases rapidly, while the relative intensities of A and B structures remain almost unchanged. For $x<1.6$, the π* contribution (C) is negligible and the spectral weight is centred around A and B with an inversion of their intensity ratio. This can be explained considering the behaviour of the cross-sections and the number of V *3d* and O *2p* electrons. The only exception to the scenario described above is represented by the VB spectrum of the sub-oxide sample, which has its maximum between A and B and where none of the features so far described can be recognized. This intermediate feature could be a superposition of A and B, *i.e.* indicate a mixed *4s-3d* valence band. Further experimental investigation and theoretical support is however necessary to clarify this issue.

Because of overlapping contributions from multiple orbitals,[45] a quantitative analysis correlating VB spectral features to stoichiometry is difficult. The fit of the spectral features with fixed line-shapes does not provide reliable value and their interpretation is questionable without a priori knowledge of the superposition degree of the contributing orbitals. To overcome this problem, we introduced a semi-empirical formula, which correlates the branching ratio of V *3d* and O *2p* bands with the stoichiometry. Starting from the assumption that a VB spectrum can be divided into its fundamental contributions, and considering V*3d*-O*2p* hybridization, we arrived at a similar equation for the outer valence branching ratio as for the Auger (eq.1).

$$BR = \frac{I_{V3d}}{I_{O2p}+I_{V3d}} = \frac{1-0,32x}{\left(\frac{\sigma_{O2p}}{\sigma_{V3d}}+a\right)x+1+K_{sd}+b} \quad (1)$$

The detailed derivation of this BR equation is reported in the supporting information. BR experimental data have been extracted from the valence band spectra by integrating the V 3d and the O 2p regions to obtain $I_{V3d}$ and $I_{O2p}$ respectively. In Fig. 4d the BR values from the full set of samples are displayed and compared with the best-fit obtained from equation 1 (red curve) using a and b as free parameters (a= 4.5±0.7 and b= 0.5±0.6 in Fig. 4d). The good agreement among the experimental points and the fit supports the use of the proposed model for stoichiometry quantification using UPS and once more it underlines the failure of the Ionic picture model to describe the valence features of vanadium oxides.

*The Work Function*

In vanadium oxides the WF increases simultaneously with the oxidation state ranging from the value of 4.3 eV of the pure vanadium up to ~7 eV as in stoichiometric $V_2O_5$.[10]

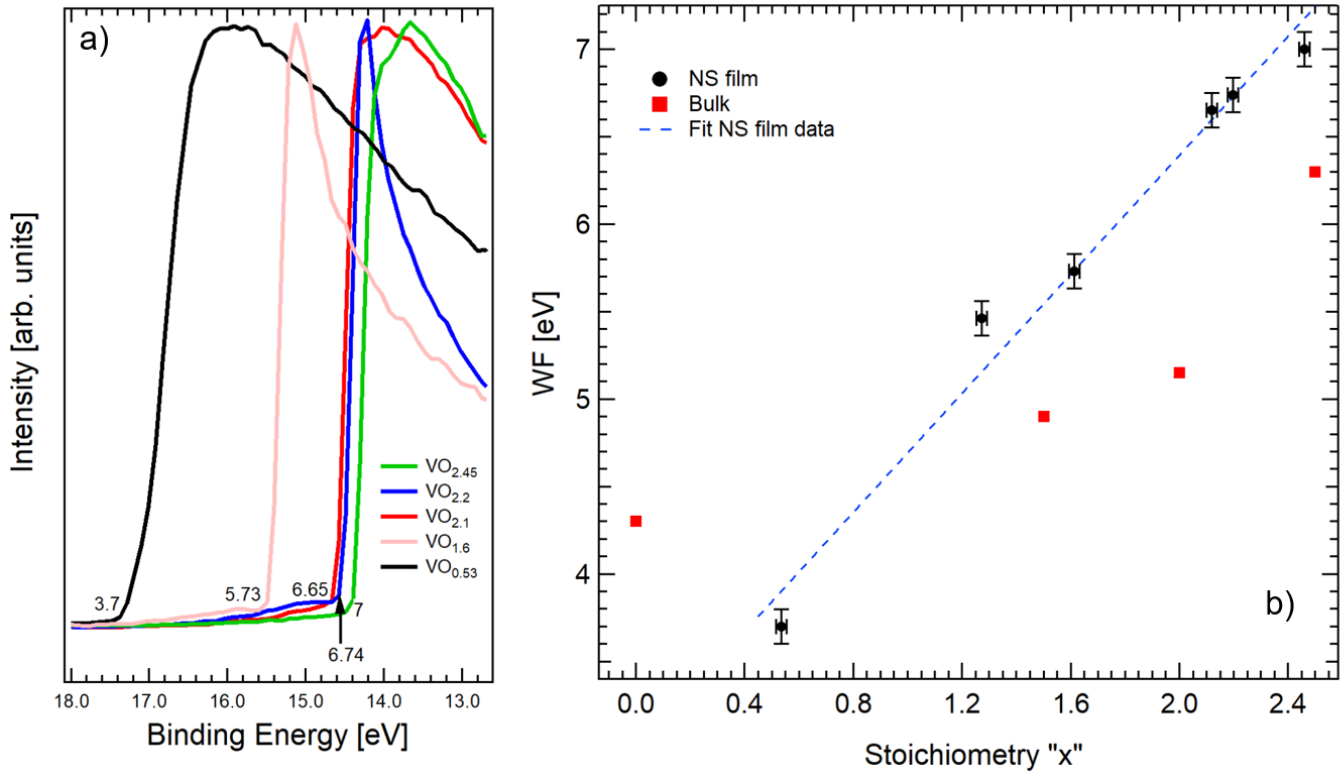

**Figure 5:** a): comparison of the onset energy of the secondary electron onset collected with the He I (21.22 eV) discharge lamp of $VO_{0.53}$, $VO_{1.6}$, $VO_{2.1}$, $VO_{2.2}$, and $VO_{2.45}$ NS films. For each film the relative WFs are also reported. b) Comparison of the WF of $VO_x$ NS films (circles) with bulk data taken from literature [6,10,13,46] (squares) for the stoichiometric ratio x in the range 0 -2.5. The dashed line is the best least square fit of the NS films data.

The measurements of the WF of these films have been performed measuring the VB spectra (see Fig. 5a) upon negatively biasing the sample by ~ -9 V to determine the secondary electron onset energy. The WF has been calculated as:

$WF = h\nu_{He\ I}$ (21.22 eV) - $BE_{So}$

where $BE_{So}$ is the binding energy corresponding to the secondary electron onset. The dependence of WF from stoichiometry in these NS vanadium oxides films is depicted in Fig. 5b.

A straight line can fit data of NS films:

$$WF = w_0 + w_1 x \qquad (2)$$

Where $w_0$ and $w_1$ are fit parameters. The slope has a value of $w_1 = 1.7 \pm 0.13\ eV$ and the intercept $w_0 = 2.99 \pm 0.23\ eV$.

Some experimental and extrapolated WF value are listed in Table 2, together with data available in the literature for bulk samples. Extrapolated and measured values of the WF of NS films are systematically larger than their bulk counterpart. This may be due to a size effect, which determines the enhancement of the electronic density by confinement within the nanostructure. It is well known since 80's that the WF of small spherical nanoparticles is larger than that of bulk.[15–18]

Table 2: WF values of NS films and reference bulk materials taken from the literature.[6,13,14,46] The extrapolated values are labelled by *. The NP diameters have been obtained using the eq. 3.

| Oxide | NS film WF [eV] | Bulk WF [eV] | NP diameter [nm] |
|---|---|---|---|
| $VO_{2.45}$ | 7±0.1 | 6.3[13] | 1.54±1.0 |
| VO$_{2.2}$ | 6.74±0.1 | - | - |
| VO$_{2.1}$ | 6.65±0.1 | - | - |
| VO$_2$ | *6.39±0.33 | 5.15[46] | 0.9±0.3 |
| VO$_{1.6}$ | 5.73±0.1 | 4.65[14] | 0.9±0.3 |
| V$_2$O$_3$ | *5.61±0.29 | 4.9[6] | 1.7±1.1 |
| VO$_{1.27}$ | 5.46±0.1 | - | - |
| VO$_{0.53}$ | 3.7±0.1 | - | - |

Zhou and Zachariah deeply investigated the size dependence of the WF of aggregated NP.[17] They claimed that for assembled NP, the deviation respect to the bulk strongly depends on the size of the primary particles and weakly from the size of the aggregate. Assuming that the fundamental units of the nanoparticle-aggregate are spherically shaped, using the eq. (3) it is possible to estimate the average diameter of each element:

$$WF_{np} = WF_{bulk} + \frac{1.08}{D_{np}} \quad (3)$$

WF is expressed in eV and the diameter ($D_{np}$) in nm. Eq. (3) cannot provide an accurate size evaluation since the spherical shape approximation is too general and is not adequate to describe the complex shape-size relation. Indeed, a fractal-like nanostructured distribution is showed by TEM (see Fig. 1 and supplementary materials) in our samples. Still, the results of eq.(3) can be indicative of the order of magnitude of the nanoparticles size. The diameters listed in Table 2, suggest that the main components have a size of the order of 1-2 nm, in agreement with the few nm particles observed in TEM images. In addition, the calculated diameters for different stoichiometries are similar, within the uncertainty. The result suggests that the nanoparticles nucleation conditions within the PMCS are not heavily affected by the small amount of oxygen present in the gas mixture, i.e. the size and shape are not depending by nanoparticles stoichiometry, as confirmed by TEM. The observation of such size effect in WF measurements, is an addtitional confirmation that that NS films synthetized with the SCBD grow without coalescence retaining the memory of the small size of their gas-phase precursors.[47–51] As a consequence of this "memory effect" SCBD allows to exploit quantum confinement to synthetize samples with a remarkably high WF and a low stoichiometry without any sputtering process or additional heating treatment.

An interesting case is that of suboxides. Despite the quantum confinement enhancement, for x<0.76 the WF of the films is smaller compared to the one of pure bulk vanadium, most likely owing to the arrangements of vanadium and oxygen atoms in suboxides that determine a thin electron depletion layer on the surface of the sample. As demonstrated by Leung and co-workers, WF modifications induced by electronegative elements depends from the fine details of the electron density at the surface, which may lead to a decrease of the work function of metal oxides.[52]

The low WF of suboxides could be extremely useful if they are used as electron injection layer in an OSD, since as in OLED the optimization of the cathode WF may increase the efficiency.[7,53] We would like to point out that the straight line trend of eq. 3 is not valid for x=0. Vanadium bulk WF is 4.3 eV, therefore, because of the above mentioned considerations on quantum confinement, for nanostructured vanadium we expect a WF>4.3 eV. This implies that exists a point x* for which the WF will start to increase, reducing the stoichiometric ratio thus disobeying the linear relation of eq. 3.

## *Conclusions*

By using the SCBD approach at room temperature, we grew different NS vanadium oxides films with controllable oxidation state. We systematically investigated using *in-situ* Auger spectroscopy $VO_x$ electronic structures as a function of the stoichiometry. From the analysis of the Auger spectra we quantitatively correlated the *3d* spectral weight of the DOS with the amount of oxygen inside the films. Results confirm the failure of the ionic picture to describe the *3d* occupancy in NS vanadium oxides, and the role of the hybridization in all VOx NS film. We have shown that the amount of unpaired *d* electron exhibits a linear dependence from the oxygen content, indicating a strong *4s-3d* and *3d-2p* hybridization varying with stoichiometry. Based on the data we propose heuristic model which links the outer valence BR with the stoichiometry, again pointing out the importance of V3d-O*2p* hybridization in these VO$_x$ films. In addition we observed the occurrence of a linear correlation between WF and stoichiometry in the range 0.5<x<2.45. The observed "memory effect" allowed to exploit the nanograin structure of these films resulting in a quantum confinement enhanced WF. By controlling the oxidation state degree of freedom and accessing the nanoscale we demonstrate the unique capability of this method to synthetize NS films with a tunable WF (3.7-7 eV), together with the possibility to grow at room temperature hole or electron injection layers. SCBD emerges as an effective approach for the control of the electronic structure and work function of materials.

*Experimental methods*

All samples were produced in-situ under ultra-high vacuum (UHV) conditions (base pressure <2×10$^{-9}$ mbar) by using the SCBD apparatus equipped with a PMCS which is available at Lab. TASC-Analytical Division[54]. The PMCS is a pulsed-cluster source driven by a high-power pulsed electric discharge. In the present experiment, the PMCS was operated with a vanadium cathode (6 mm diam. rod, purity 99.9 %, EvoChem GmbH) generating a supersonic beam of vanadium metal or oxide cluster. To obtain homogeneously oxidized nanoparticles, we used Ar (high purity Ar: 99.9995%, SIAD) as carrier gas, mixed with a controlled amount of oxygen (Table 1, supplementary materials) resulting in an Ar-O$_2$ gas mixture. The working parameters of the PMCS have been kept constant for all the samples synthesized (delay between gas injection and discharge firing = 0.6 ms; discharge operating voltage 0.925 kV; discharge duration 80 µs; pulsed-valve aperture driving signal duration time 157 µs; pulse repetition rate 3 Hz; Ar pressure 70 bar). The nanostructured film deposition rate measured by a quartz-crystal microbalance ranged from ~30 Å/s for pure Ar as carrier gas, to ~5 Å /s for the highest oxygen concentration. To probe the vanadium oxidation state, we deposited VOx films with >300 nm thickness on Si or Cu substrates. The typical size of all the samples is about 2 cm$^2$. The samples are deposited by landing the supersonic beam of nanoparticles onto clean substrates placed inside the sample-preparation chamber of the XPS characterization facility (base pressure <2×10-9 mbar) and then transferred, maintaining UHV conditions, via a linear translation manipulator into the analysis chamber where pressure was kept below 5×10-10 mbar. Here they are characterized by X-ray Photoemission Spectroscopy (XPS) using a Mg Kα lamp (1253.6 eV, not monochromatic) and studied by UPS using a He lamp (He I 21.22 eV) coupled to a hemispherical analyser (PSP, 120 mm). The morphology and the size of the clusters has been analysed by Transmission Electron Microscopy (TEM) by means of a field-emission TEM (JEM2100F-UHR, JEOL) operated with beam energy 200 KV.

deposition: An X-ray Photoelectron Spectroscopy characterization. *Thin Solid Films* **520**, 4803–4807 (2012).

## Additional information:

**Acknowledgments:** The authors would like to acknowledge F. Zuccaro (ISM-CNR) for the technical support and the staff of Low Density Matter beamline @ FERMI for assistance in preparing the gas mixture. Dr. N. Zema (ISM-CNR), Dr. C. Spezzani (Elettra) and Dr. M. Sacchi (CNRS) for the support with the XANES measurements.

**Author Contributions**: Conceptualization, A.D., P.P., M.C. and A.M.; methodology A.D., C.C. and M.C.; investigation, M.F., B.B., C.C., A.D. and P.P.; data curation A.D., B.B. and S.M.; resources M.C., C.C. and M. d. S.; writing-original draft preparation, A.D.; all the authors reviewed the manuscript.

**Competing Interests**: The authors declare no competing interests.

**Supporting Information Available**: Table S1-S3 and Figure S1-S12.

## Figures legends and tables:

**Figure 1:** TEM image of the nanoparticle assembled nanostructured film for $VO_{2.2}$

**Figure 2:** V2p and O1s spectra of vanadium and oxygen. a) V2p and O1s core level spectra of all samples. Peak positions of the vanadium components for different oxidation state are highlighted by straight vertical lines; b) V2p and O1s core level spectra and fit curve of $VO_{2.2}$. The individual fit components are shown as coloured curves. The label O 1s sat indicates the features generated by the Mg K$\alpha_3$ and Mg K$\alpha_4$ component of the radiation.

**Figure 3:** a) Comparison of Auger L3M2,3M4,5 spectra of vanadium oxides NS films with stoichiometric ratio 1.27, 1.6 and 2.2 in the kinetic energy range 455-482 eV. The spectra are vertically shifted for clarity in order to highlight the behavior of features labelled U and V (see text). b) Correlation between stoichiometry and Auger fraction (U/(U+V)) as obtained by fitting the Auger L-$_3$M$_{2,3}$M$_{4,5}$ spectra. The blue line is the best least-squares fit of the experimental data with a straight line of fixed intercept (at x=0 the fraction of unpaired electrons must be equal to 1). The best-fit slope is k = -0.320±0.005

**Figure 4:** a) full valence band spectrum of VO$_{2.45}$ in the BE region [0-10] eV reported as representative spectra. O 2p and V 3d regions of the valence band spectra (b and c respectively) for VO$_{0.53}$, VO$_{1.27}$, VO$_{1.6}$, VO$_{2.1}$, VO$_{2.2}$ and VO$_{2.45}$ oxides: b) Comparison of the spectra of the O 2p region. Spectra are vertically shifted in order to highlight the behavior of the feature labelled A, B and C; c) comparison of the spectra in the V 3d region, near the Fermi level (dotted line). The arrows indicate VO$_{1.6}$ features. The intensity of the spectra is normalized to the maximum value of O2p region. d) BR as a function of the stoichiometric ratio. The red curve is the fit obtained using the eq. 1 including the 3d-2p hybridization term; the blue dotted curve is obtained considering ionic model of the valence band.

**Figure 5:** a): comparison of the onset energy of the secondary electron onset collected with the He I (21.22 eV) discharge lamp of VO$_{0.53}$, VO$_{1.6}$, VO$_{2.1}$, VO$_{2.2}$, and VO$_{2.45}$ NS films. For each film the relative WFs are also reported. b) Comparison of the WF of VO$_x$ NS films (circles) with bulk data taken from literature [6,10,13,46] (squares) for the stoichiometric ratio x in the range 0 -2.5. The dashed line is the best least square fit of the NS films data.